\documentclass[aps, pra, twocolumn,showpacs,preprintnumbers,amsmath,amssymb]{revtex4}
\newcommand{\ket}[1]{\mbox{$ | #1 \rangle $}}

\usepackage[dvips]{graphicx}% Include figure files

\begin{document}

\preprint{}

\title{Quantum secure direct communication without using perfect quantum channel}
% Force line breaks with \\

\author{Jian Wang}

 \email{jwang@nudt.edu.cn}

\affiliation{School of Electronic Science and Engineering,
\\National University of Defense Technology, Changsha, 410073, China }
\author{Quan Zhang}
\affiliation{School of Electronic Science and Engineering,
\\National University of Defense Technology, Changsha, 410073, China }
\author{Chao-jing Tang}
\affiliation{School of Electronic Science and Engineering,
\\National University of Defense Technology, Changsha, 410073, China }

 %\email{ishizuka@isl.melco.co.jp}
% \affiliation{School of Electronic Science and Engineering, \\National Univ of
%Defense Technology, Changsha , 410073,China}
%{Information Technology R \& D Center, Mitsubishi Electric
%Corporation\\
%5-1-1 Ofuna, Kamakura, Kanagawa 247-8501, JAPAN\\
%TEL: +81-467-41-2190 \quad FAX: +81-467-41-2185}


\begin{abstract}
Most of the quantum secure direct communication protocol needs a
pre-established secure quantum channel. Only after insuring the
security of quantum channel, could the sender encode the secret
message and send them to the receiver through the secure channel. In
this paper, we present a quantum secure direct communication
protocol using Einstein-Podolsky-Rosen pairs. It is not necessary
for the present protocol to insure the security of quantum channel
before transmitting the secret message. In the present protocol, all
Einstein-Podolsky-Rosen pairs are used to transmit the secret
message except those chosen for eavesdropping check.

\end{abstract}

%Use showkeys class option if keyword
\maketitle

%%%%%%%%%%%%%%%%%%%%%%%%%%%%%%%%%%%%%%%%%%%%%%%%%%%%%%%%%%%%%%%%%%%%%
%=====================================================================
\section{Introduction}
%=====================================================================
%
%
Quantum key distribution (QKD) is one of the most promising
applications of quantum information science. The goal of QKD is to
allow two legitimate parties, Alice and Bob, to generate a secret
key over a long distance, in the presence of an eavesdropper, Eve,
who interferes with the signals. Since Bennett and Brassard
presented the original QKD protocol \cite{bb84}, it has been
developed quickly. Recently, a novel concept, quantum secure direct
communication (QSDC) has been proposed \cite{beige}. Different from
QKD whose object is to establish a common key between the
communication parties, QSDC's object is to transmit the secret
messages directly without first establishing a key to encrypt them.
In this paper, we follow the above definition of QSDC.

QSDC can be used in some special environments which has been shown
by Bostr\"{o}em and Deng et al. \cite{Bostrom,Deng}. The works on
QSDC attracted a great deal of attentions [2-8]. We can divide these
works into two kinds, one utilizes entangled state
\cite{Bostrom,Deng,yan,gao,wang}, the other utilizes sing-photon
\cite{denglong}. Bostr\"{o}m and Felbinger proposed a Ping-Pong QSDC
protocol which is quasi-secure for secure direct communication if
perfect quantum channel is used \cite{Bostrom}. Deng et al. put
forward a two-step QSDC protocol using Einstein-Podolsky-Rosen (EPR)
pairs (hereafter called Deng's protocol) \cite{Deng} and a QSDC
scheme with a quantum one-time pad \cite{denglong}. Feng Li Yan and
Xiao Qiang Zhang presented a QSDC scheme using EPR pairs and
teleportation (hereafter called Yan's scheme) \cite{yan}. The
security of most of QSDC schemes relies on a pre-established secure
quantum channel. Deng et al. pointed out the basic requirement for
QSDC scheme in \cite{denglong}, which eavesdropping check before the
message being encoded must be performed first. Only in this way, can
the sender utilize the checked channel to encode the secret message
and transmit it to the receiver.

Actually, the communication parties, Alice and Bob can obtain a
common key as long as a secure quantum channel is established. Alice
can then send her secret messages to Bob directly by classical
channel. For example, in Yan's scheme \cite{yan}, Alice and Bob can
obtain a common key as long as they measure their particles in the
$Z$-basis. It is not necessary for the sender to transmit her secret
to the receiver using teleportation after insuring the security of
quantum channel. Note that it needs two-bit classical information to
recovery one bit secret message in Yan's scheme. It would be better
to transmit directly the secret to the receiver using classical
channel. Suppose the secret messages to be transmitted is 101001 and
the common key between Alice and Bob is 001101. Alice tells Bob to
reorder his results of measurements by choosing his result of the
third, the first, the fourth, the second, the fifth, and the sixth
particle in turn. That is to say Bob can recover Alice's secret
messages 101001 as long as he reorders his results of measurements
according Alice's classical messages and the eavesdropper, Eve
cannot obtain any information about the secret.

In this paper, we present a QSDC protocol with EPR pairs without
insuring the security of quantum channel before transmitting the
secret message. The secret message is deterministically sent through
the quantum channel. We also show the present protocol is secure and
efficient.

This paper is organized as follows. In Sec.\ref{protocol}, we
describe the process of the QSDC protocol. In Sec.\ref{security}, we
discuss the security and efficiency of the present protocol.
Finally, we give a summary in Sec.\ref{conclusion}.

%%%%%%%%%%%%%%%%%%%%%%%%%%%%%%%%%%%%%%%%%%%%%%%%%%%%%%%%%%%%%%%%%%%%%
%=====================================================================
\section{The QSDC Protocol}
\label{protocol}
%=====================================================================
In the QSDC protocol, we suppose the sender Alice wants to send a
secret message to the receiver, say Bob directly. The basic idea of
the protocol originates from quantum teleportation\cite{bbc93}.
Alice entangles his encoded secret message state with prepared EPR
pairs. She then performs controlled-NOT (CNOT) operation and
Hadamard transformation, which is similar to the method used in
quantum teleportation. Yan's scheme utilized the method of quantum
teleportation faithfully. After insuring the security of quantum
channel, the receiver performs unitary operation on his particle to
recover the sender's secret message according to the sender's
classical message. The basic idea of our protocol is different from
Yan's scheme. Instead of performing unitary operation to recover
Alice's secret message, Bob measures his particle in a fixed
measuring basis and recovers Alice's secret message according to the
correlation between two parties' results. The QSDC protocol is as
follows:

(1)Alice prepares an ordered $N$ EPR pairs in the state
\begin{eqnarray}
\ket{\phi}_{AB}=\frac{1}{\sqrt{2}}(\ket{00}+\ket{11})_{AB}.
\end{eqnarray}
We denotes the ordered $N$ EPR pairs with \{[P$_1(A)$,P$_1(B)$],
[P$_2(A)$,P$_2(B)$], $\cdots$, [P$_N(A)$,P$_N(B)$]\}, where the
subscript indicates the pair order in the sequence, and $A$, $B$
represents the two particles of each EPR pair, respectively. Alice
takes one particle from each EPR pair to form an ordered EPR partner
particle sequence [P$_1(A)$, P$_2(A)$,$\cdots$, P$_N(A)$], called
$A$ sequence. The remaining EPR partner particles compose $B$
sequence, [P$_1(B)$, P$_2(B)$,$\cdots$, P$_N(B)$]. Alice transmits
$B$ sequence to Bob.

(2) Bob selects randomly a sufficiently large subset of $B$ sequence
and performs Hadamard transformations
\begin{eqnarray}
H=\frac{1}{\sqrt{2}} \left( \begin{array}{c c} 1 & 1 \\
1 & -1
\end{array} \right)
\end{eqnarray}
on them. He then announces publicly the position of the selected
particles. The Hadamard transformation is crucial for the security
of the protocol as we will see in the sequel.

(3) After hearing from Bob, Alice executes Hadamard transformations
on the corresponding particles of $A$ sequence. She then selects
randomly a sufficiently large subset of particles from $A$ sequence,
which we call $C$ sequence. Alice generates a random bit string and
encodes it on $C$ sequence. If Alice's random bit is ``0''(``1''),
she prepares a particle $a$ in the state
$\ket{+}=\frac{1}{\sqrt{2}}(\ket{0}+\ket{1})$
($\ket{-}=\frac{1}{\sqrt{2}}(\ket{0}-\ket{1})$) for each particle of
$C$ sequence. $C$ sequence is used to check eavesdropping, which we
call checking sequence. The remaining particles of $A$ sequence
forms $D$ sequence. $D$ sequence is used to encode Alice's secret
message, which we call encoding sequence. Alice then encodes her
secret message on $D$ sequence. If Alice's secret message is
``0''(``1''), she prepares a particle $a$ in the state $\ket{+}$
($\ket{-}$) for each particle of $D$ sequence. Thus Alice prepares
$N$ particles for each particle of $A$ sequence, which we call $a$
sequence [P$_1(a)$, P$_2(a)$,$\cdots$, P$_N(a)$].

(4) If the state of the particle P$_i(a)$ is $\ket{+}$, then the
state of the particle P$_i(a)$, P$_i(A)$, and P$_i(B)$
($i=1,2,\cdots,N$) is
\begin{eqnarray}
\ket{\Phi_0}_{aAB}=\frac{1}{\sqrt{2}}(\ket{0}+\ket{1})_a\otimes\frac{1}{\sqrt{2}}(\ket{00}+\ket{11})_{AB},
\end{eqnarray}
where the subscript $a$ denotes the particle P$_i(a)$. If the state
of the particle P$_i(a)$ is $\ket{-}$, then the state of the
particle P$_i(a)$, P$_i(A)$, and P$_i(B)$ is
\begin{eqnarray}
\ket{\Phi_1}_{aAB}=\frac{1}{\sqrt{2}}(\ket{0}-\ket{1})_a\otimes\frac{1}{\sqrt{2}}(\ket{00}+\ket{11})_{AB}.
\end{eqnarray}

(5) Alice sends the particle P$_i(a)$, P$_i(A)$ ($i=1,2,\cdots,N$)
through a CNOT gate (the particle P$_i(a)$ is the controller, the
particle P$_i(A)$ is the target). Then $\ket{\Phi_0}_{aAB}$ is
changed to
\begin{eqnarray}
\ket{\Phi_0'}_{aAB}=\frac{1}{2}(\ket{000}+\ket{110}+\ket{011}+\ket{101})_{aAB},
\end{eqnarray}
and $\ket{\Phi_1}_{aAB}$ becomes
\begin{eqnarray}
\ket{\Phi_1'}_{aAB}=\frac{1}{2}(\ket{000}-\ket{1100}+\ket{0111}-\ket{101})_{aAB}.
\end{eqnarray}

(6) Alice performs Hadamard transformation on the particle P$_i(a)$,
obtaining
\begin{eqnarray}
\label{7} \ket{\Phi_0''}_{aAB}&=&\frac{1}{2}[\ket{00}_{aA}\otimes\frac{1}{\sqrt{2}}(\ket{0}+\ket{1})_{B}\nonumber\\
& &+\ket{10}_{aA}\otimes\frac{1}{\sqrt{2}}(\ket{0}-\ket{1})_{B}\nonumber\\
& &+\ket{01}_{aA}\otimes\frac{1}{\sqrt{2}}(\ket{0}+\ket{1})_{B}\nonumber\\
& &+\ket{11}_{aA}\otimes\frac{1}{\sqrt{2}}(\ket{1}-\ket{0})_{B}]
\end{eqnarray}
or
\begin{eqnarray}
\label{8} \ket{\Phi_1''}_{aAB}&=&\frac{1}{2}[\ket{00}_{aA}\otimes\frac{1}{\sqrt{2}}(\ket{0}-\ket{1})_{B}\nonumber\\
& &+\ket{10}_{aA}\otimes\frac{1}{\sqrt{2}}(\ket{0}+\ket{1})_{B}\nonumber\\
& &+\ket{01}_{aA}\otimes\frac{1}{\sqrt{2}}(\ket{1}-\ket{0})_{B}\nonumber\\
& &+\ket{11}_{aA}\otimes\frac{1}{\sqrt{2}}(\ket{0}+\ket{1})_{B}].
\end{eqnarray}

(7) Alice then measures the particle P$_i(a)$, P$_i(A)$ in the
$Z$-basis, $\{\ket{0}, \ket{1}\}$. Bob measures the particle
P$_i(B)$ in the $X$-basis, \{\ket{+}, \ket{-}\}. At this step,
although Bob obtains his result of measurement, he cannot recover
Alice's secret message without Alice's result. We can draw the above
conclusion according to the equation \ref{7} and \ref{8}.

(8) Alice informs Bob the positions of $C$ sequence (checking
sequence) and lets him announce his corresponding results of
measurements. Alice judges whether her random bits can be
reconstructed correctly by combining Bob's results and her results
of $C$ sequence. If the error rate is small, Alice can conclude that
there is no eavesdroppers in the line. Alice and Bob continue to
perform the next step, otherwise they abort the communication.

(9) If Alice is certain that there is no eavesdropping, she
announces the results of measurements of $D$ sequence. Thus Bob can
recover Alice's secret message, according to Alice's result, as
illustrated in Table 1.

\begin{table}[h]
\caption{The recovery of Alice's secret message }\label{Tab:one}
  \centering
    \begin{tabular}[b]{|c|c|c| c|} \hline
      Alice's result & Bob's result & secret message\\ \hline
      \ 0 & \ket{+} & 0\\ \hline
       \ 0 & \ket{-} & 1\\ \hline
        \ 1 & \ket{+} & 1\\ \hline
         \ 1 & \ket{-} & 0\\ \hline
        \end{tabular}
\end{table}
Suppose Bob's result is $\ket{+}$. If Alice's result of measurement
of the corresponding $D$ sequence particle is ``0'' (``1''), they
then conclude that the Alice's secret message is ``0'' (``1'').

%%%%%%%%%%%%%%%%%%%%%%%%%%%%%%%%%%%%%%%%%%%%%%%%%%%%%%%%%%%%%%%%%%%%%
%=====================================================================
\section{Security and efficiency of the QSDC protocol}
\label{security}
%=====================================================================

So far we have proposed the QSDC protocol. We now discuss the
security of the present protocol. The crucial point is that the
Hadamard gate at the step 2 and 3 of the scheme do not allow an
eavesdropper, Eve to have a successful attack and Eve's attack will
be detected during the eavesdropping check.

We first consider the intercept-resend attack strategy. In this
attack strategy, Eve intercepts the particles of $B$ sequence
transmitted to Bob and makes measurements on them. Then she resends
a particle sequence to Bob according to her results of measurements.
Eve can only intercept $B$ sequence at the step 1 of the protocol
and she cannot make certain which particle will be executed Hadamard
transformation. Suppose Eve measures the intercepted particle on
which Alice and Bob will not perform Hadamard transformation in the
$Z$-basis. In this way, if the result of Eve's measurement is ``0'',
she sends a particle in the state $\ket{+}$ to Bob, otherwise sends
a particle in the state $\ket{-}$. Then if Alice prepares a
particle, P$_i(a)$ in the state $\ket{+}$, the state of [P$_i(a)$,
P$_i(A)$, P$_i(B)$] collapses to $\ket{+0+}_{aAB}$ or
$\ket{+1-}_{aAB}$ each with probability 1/2. After Alice's CNOT
operation and Hadamard transformation, the state becomes
\begin{eqnarray}
\ket{\Psi_i}_{aAB}=\frac{1}{2}(\ket{00}+\ket{01}+\ket{10}-\ket{11})_{aA}\otimes\ket{+}_B
\end{eqnarray}
or
\begin{eqnarray}
\ket{\Psi_i}_{aAB}=\frac{1}{2}(\ket{00}+\ket{01}-\ket{10}+\ket{11})_{aA}\otimes\ket{-}_B.
\end{eqnarray}
Alice measures P$_i(a)$ in the $Z$-basis and obtains ``0'' or ``1'',
each with probability 1/2. She has only 50\% probability of
obtaining the right result. During the eavesdropping check, Eve's
attack will be detected easily. Similarly, If Alice prepares a
particle, P$_i(a)$ in the state $\ket{-}$, the error rate introduced
by Eve will also achieve 50\%.

Suppose Eve measures the intercepted particle on which Alice and Bob
will perform Hadamard transformation in the $X$-basis. Note that
$\ket{\phi}_{AB}$ can also be expressed as
$\frac{1}{\sqrt{2}}(\ket{++}+\ket{--})_{AB}$. If Alice prepares a
particle, P$_i(a)$ in the state $\ket{+}$, then The state of
[P$_i(a)$, P$_i(A)$, P$_i(B)$] collapses to $\ket{+++}_{aAB}$ or
$\ket{+--}_{aAB}$ each with probability 1/2. After the Hadamard
transformations of Alice and Bob, the state becomes
$\ket{+00}_{aAB}$ or $\ket{+11}_{aAB}$. According to the protocol,
the state is changed to
\begin{eqnarray}
\ket{\Psi_i'}_{aAB}=\frac{1}{2\sqrt{2}}(\ket{0+}+\ket{1-})_{aA}\otimes(\ket{+}+\ket{-})_B
\end{eqnarray}
or
\begin{eqnarray}
\ket{\Psi_i'}_{aAB}=\frac{1}{2\sqrt{2}}(\ket{0+}-\ket{1-})_{aA}\otimes(\ket{+}-\ket{-})_B
\end{eqnarray}
Obviously, Eve's eavesdropping will be detected during the
eavesdropping check. It will have the same result if Alice prepares
a particle, P$_i(a)$ in the state $\ket{-}$.

We then consider the collective attack strategy. In this strategy,
Eve intercepts the particle P$_i(B)$ and uses it and her own
ancillary particle in the state $\ket{0}$ to do a CNOT operation
(the particle P$_i(B)$ is the controller, Eve's ancillary particle
is the target). Then Eve resends the particle P$_i(B)$ to Bob.
However, Eve cannot make certain which particle will be performed
Hadamard transformation. Suppose Bob will not perform Hadamard
transformation on the intercepted particle. The state of P$_i(A)$,
P$_i(B)$, and Eve's corresponding ancillary particle is
\begin{eqnarray}
\ket{\Omega}_{ABE}=\frac{1}{\sqrt{2}}(\ket{000}+\ket{111})_{ABE},
\end{eqnarray}
where the subscript $E$ indicates Eve's ancillary particle.
According to the protocol, the state of the particle P$_i(a)$,
P$_i(A)$, P$_i(B)$, and the corresponding Eve's ancillary particle
will be
\begin{eqnarray}
\ket{\Omega_0}_{aABE}&=&\frac{1}{2}[\ket{00}_{aA}\otimes\frac{1}{\sqrt{2}}(\ket{00}+\ket{11})_{BE}\nonumber\\
& &+\ket{10}_{aA}\otimes\frac{1}{\sqrt{2}}(\ket{00}-\ket{11})_{BE}\nonumber\\
& &+\ket{01}_{aA}\otimes\frac{1}{\sqrt{2}}(\ket{00}+\ket{11})_{BE}\nonumber\\
& &+\ket{11}_{aA}\otimes\frac{1}{\sqrt{2}}(\ket{11}-\ket{00})_{BE}]
\end{eqnarray}
or
\begin{eqnarray}
\ket{\Omega_1}_{aABE}&=&\frac{1}{2}[\ket{00}_{aA}\otimes\frac{1}{\sqrt{2}}(\ket{00}-\ket{11})_{BE}\nonumber\\
& &+\ket{10}_{aA}\otimes\frac{1}{\sqrt{2}}(\ket{00}+\ket{11})_{BE}\nonumber\\
& &+\ket{01}_{aA}\otimes\frac{1}{\sqrt{2}}(\ket{11}-\ket{00})_{BE}\nonumber\\
& &+\ket{11}_{aA}\otimes\frac{1}{\sqrt{2}}(\ket{00}+\ket{11})_{BE}].
\end{eqnarray}
Note that
\begin{eqnarray}
\frac{1}{\sqrt{2}}(\ket{00}+\ket{11})=\frac{1}{\sqrt{2}}(\ket{++}+\ket{--})
\end{eqnarray}
and
\begin{eqnarray}
\frac{1}{\sqrt{2}}(\ket{00}-\ket{11})=\frac{1}{\sqrt{2}}(\ket{+-}+\ket{-+}).
\end{eqnarray}
We can rewritten $\ket{\Omega_0}_{aABE}$ and $\ket{\Omega_1}_{aABE}$
as
\begin{eqnarray}
\frac{1}{2}[\ket{00}_{aA}\otimes\frac{1}{\sqrt{2}}(\ket{++}+\ket{--})_{BE}\nonumber\\
+\ket{10}_{aA}\otimes\frac{1}{\sqrt{2}}(\ket{+-}-\ket{-+})_{BE}\nonumber\\
+\ket{01}_{aA}\otimes\frac{1}{\sqrt{2}}(\ket{++}+\ket{--})_{BE}\nonumber\\
-\ket{11}_{aA}\otimes\frac{1}{\sqrt{2}}(\ket{+-}+\ket{-+})_{BE}]
\end{eqnarray}
and
\begin{eqnarray}
\frac{1}{2}[\ket{00}_{aA}\otimes\frac{1}{\sqrt{2}}(\ket{+-}+\ket{-+})_{BE}\nonumber\\
+\ket{10}_{aA}\otimes\frac{1}{\sqrt{2}}(\ket{++}-\ket{--})_{BE}\nonumber\\
-\ket{01}_{aA}\otimes\frac{1}{\sqrt{2}}(\ket{+-}+\ket{-+})_{BE}\nonumber\\
+\ket{11}_{aA}\otimes\frac{1}{\sqrt{2}}(\ket{++}+\ket{--})_{BE}]
\end{eqnarray}
Alice then measures the particle P$_i(a)$ in the $Z$-basis and Bob
measures the particle P$_i(B)$ in the $X$-basis. Suppose Alice's
random bit is ``0'' and her result of measurement of the
corresponding $C$ sequence particle is ``0'' (``1''). According to
the protocol, Bob's result must be $\ket{+}$ ($\ket{-}$). Similarly,
if Alice's random bit is ``1'' and her result of measurement of the
corresponding $C$ sequence particle is ``0'' (``1''). Thus Bob's
result must be $\ket{-}$ ($\ket{+}$). However, Bob can only obtains
the right result with probability 1/2 because of Eve's
eavesdropping. During the eavesdropping check, half of Bob's results
will be inconsistent with that of Alice's. Thus Eve's eavesdropping
will be detected easily, because her eavesdropping introduces a
error rate with 50\%. Eve measures the intercepted particle, but she
can only obtain $\ket{+}$ or $\ket{-}$ each with probability 1/2.
Because she has no information about Alice's result, she cannot
conclude what Alice's secret message is, even if she obtains Bob's
result of measurement.

Suppose Eve execute Hadamard and CNOT operation on the intercepted
particle which Bob will perform Hadamard transformation on it. Note
that $\ket{\phi}_{AB}$ can be expressed as
$\frac{1}{\sqrt{2}}(\ket{++}+\ket{--})_{AB}$. Thus the state of
P$_i(A)$, P$_i(B)$ and Eve's corresponding ancillary particle will
be
\begin{eqnarray}
\ket{\Omega_{ABE}}=\frac{1}{\sqrt{2}}(\ket{++0}+\ket{--1})_{ABE}.
\end{eqnarray}
After Hadamard transformation of Alice and Bob, $\ket{\Omega_{ABE}}$
is changed to $\frac{1}{\sqrt{2}}(\ket{000}+\ket{111})_{ABE}$ which
is equal to the equation 13. As we described above, Eve's
eavesdropping can also be detected.

We now analyze the efficiency of the present protocol. In our
protocol, all EPR pairs are used to transmit the secret message
except those chosen for checking eavesdroppers because the measuring
basis of communication parties is invariable. We only need to
transmit particles once during the process of protocol, so we only
need eavesdropping check once. Twice transmission of EPR particles
and twice eavesdropping check are required in Deng's scheme.

%%%%%%%%%%%%%%%%%%%%%%%%%%%%%%%%%%%%%%%%%%%%%%%%%%%%%%%%%%%%%%%%%%%%%
%=====================================================================
\section{Summary}
\label{conclusion}
%=====================================================================
So far we have proposed a QSDC scheme using EPR pairs and analyzed
the security and efficiency of the present protocol. To prevent
eavesdropping, Alice and Bob perform Hadamard transformation on the
randomly selected particles. Alice encodes her secret message into a
given state and sends it to the receivers directly using quantum
channel. Different from most of the QSDC schemes,  it is not
necessary for our protocol to insure the security of quantum channel
before sending Alice's secret message. Without Alice's result of
measurement, the receiver or the eavesdropper cannot have any
information about Alice's secret message. Alice announces her result
only if she is certain that there is no eavesdropping in the line
and only in such a way can the receiver recover Alice's secret
message. The present protocol is efficient in that all EPR pairs are
used to transmit the secret message except those chosen for
eavesdropping check.

%%%%%%%%%%%%%%%%%%%%%%%%%%%%%%%%%%%%%%%%%%%%%%%%%%%%%%%%%%%%%%%%%%%%%%

%%%%%%%%%%%%%%%%%%%%%%%%%%%%%%%%%%%%%%%%%%%%%%%%%%%%%%%%%%%%%%%%%%%%%%

\begin{acknowledgments}
This work is supported by the National Natural Science Foundation of
China under Grant No. 60472032.
\end{acknowledgments}

%%%%%%%%%%%%%%%%%%%%%%%%%%%%%%%%%%%%%%%%%%%%%%%%%%%%%%%%%%%%%%%%%%%%%
%
%

%%%%%%%%%%%%%%%%%%%%%%%%%%%%%%%%%%%%%%%%%%%%%%%%%%%%%%%%%%%%%%%%%%%%%
%
%
\end{document}